\documentclass[a4paper]{aipproc}

\layoutstyle{6x9}
  
\begin{document}

\title 
      {Single particle properties of $\Lambda$ hypernuclei in
       the density dependent relativistic hadron field theory}


\author{Christoph Keil}{
  address={Theoretische Physik Uni Gie\ss en, Heinrich-Buff-Ring 16, 35392 Gie\ss en, Germany},
  email={christoph.m.keil@theo.physik.uni-giessen.de},
  thanks={This work was supported by DFG}
}

\iftrue
\author{Horst Lenske}{
  address={Theoretische Physik Uni Gie\ss en, Heinrich-Buff-Ring 16, 35392 Gie\ss en, Germany}
}

\fi

\copyrightholder{Christoph Keil}
\copyrightyear  {2001}

\begin{abstract}
The density dependent relativistic hadron field theory is used to describe single 
particle properties of $\Lambda$ hypernuclei. The discussion focuses on
the spin-orbit systematics in the relativistic mean-field formalism by discussing general
effects of the $\Lambda$ continuum threshold and the delocalization of the $\Lambda$ 
wave function. Theoretical predictions for the hypernuclear Auger effect
are presented.
\end{abstract}

\date{\today}

\maketitle

\section{Introduction}
\label{sec:intro}
Due to the experimental difficulties in persuing hyperonic scattering experiments
the modeling of hypernuclei is crucial for investigating interactions in the baryon octet.
For this reason the development and application of microscopic models linking
in-medium and free interactions is necessary. The extension
of the density dependent relativistic hadron field theory (DDRH) 
\cite{Keil:2000hk} into the strangeness sector and to hypernuclei is such a
microscopic link. Besides an appropriate microscopic description it is
essential to chose suitable systems in which dynamic many-body effects play a minor role
in order to get more direct access to generic hyperon interactions.
In this respect we present our studies concerned with {\em weakly bound 
$\Lambda$ states} in hypernuclei. It is found that deeply bound
$\Lambda$ states are fairly save from subtle dynamical effects -- though caution needs
to be taken there in other respects. Such deeply bound $\Lambda$ states can ideally be found
in heavy hypernuclei. For the investigation of heavy hypernuclei there
is, besides the standard spectroscopic methods, also the possibility of the spectroscopy
of {\em Auger neutrons} \cite{Likar:1986jj} as planned in a recently proposed experiment
at JLab \cite{JLabAuger}. We present Auger transition rates for $^{209}_\Lambda$Pb 
calculated within our self-consistent model and discuss the spectroscopic content of
the spectra.

\section{The DDRH model}
\label{sec:ddrh}
The DDRH model \cite{Keil:2000hk} is a relativistic Lagrangian
field theory with baryonic and mesonic degrees of freedom. In-medium interactions are
derived from a Dirac-Brueckner (DB) calculation using the free space Bonn A interaction.
The DB self-energies are mapped onto DDRH self-energies leading to vertex
functionals that depend on Lorentz scalars of the baryonic field operators.
The Lagrange density of DDRH is given by ${\cal L}={\cal L}_B+{\cal L}_M+{\cal L}_{int}$.
${\cal L}_B$ and ${\cal L}_M$ denote the free baryonic and mesonic Lagrange densities,
whereas the interaction lagrangian is given by:
\begin{eqnarray}
\mathcal{L}_{int} &=& \overline{\Psi}_F \hat{\Gamma}_\sigma(\overline{\Psi}_F, \Psi_F) \Psi_F \sigma
- \overline{\Psi}_F \hat{\Gamma}_\omega(\overline{\Psi}_F, \Psi_F) \gamma_\mu \Psi_F \omega^\mu \\
&&- \frac{1}{2}\overline{\Psi}_F \hat{\vec\Gamma}_\rho(\overline{\Psi}_F, \Psi_F) \gamma_\mu \Psi_F
\vec\rho^\mu
- e \overline{\Psi}_F \hat{Q} \gamma_\mu \Psi_F A^\mu \nonumber
\end{eqnarray}
Due to the functional structure of the vertices $\Gamma(\overline{\Psi}_F, \Psi_F)$
the baryonic equations of motion contain additional {\em rearrangement self-energies} which 
arise due to the variation of ${\cal L}_{int}$ with respect to $\Psi$. These 
rearrangement self-energies account for static polarization effects in the baryonic 
medium, assuring the thermodynamical consistency of the theory and Lorentz covariance of the
field equations \cite{Keil:2000hk}.

\section{Single particle properties of $\Lambda$ hypernuclei}

\subsection{Weakly bound $\Lambda$ states}
\label{sec:loose}
In this section we point out the subtleties of extracting 
information on generic hyperon interactions from weakly bound $\Lambda$ systems on
two examples: (1) the influence of the close continuum threshold on the structure of $\Lambda$
spin-orbit (s.o.) doublets and (2) the influence of the delocalization of the $\Lambda$
wave function resulting from the weak binding of the $\Lambda$. These effects are
{\em independent of the specific interaction} and, in general, contribute as an apparent quenching
of spin-orbit strengths.

\subsubsection{Continuum threshold level compression}
\begin{figure}
\label{fig:so}
\caption{The 1p $\Lambda$ s.o. splitting $\Delta$ E is plotted versus the centroid
         energy E$_{centroid}$. The insert defines the notation used.}
\includegraphics[height=.2\textheight]{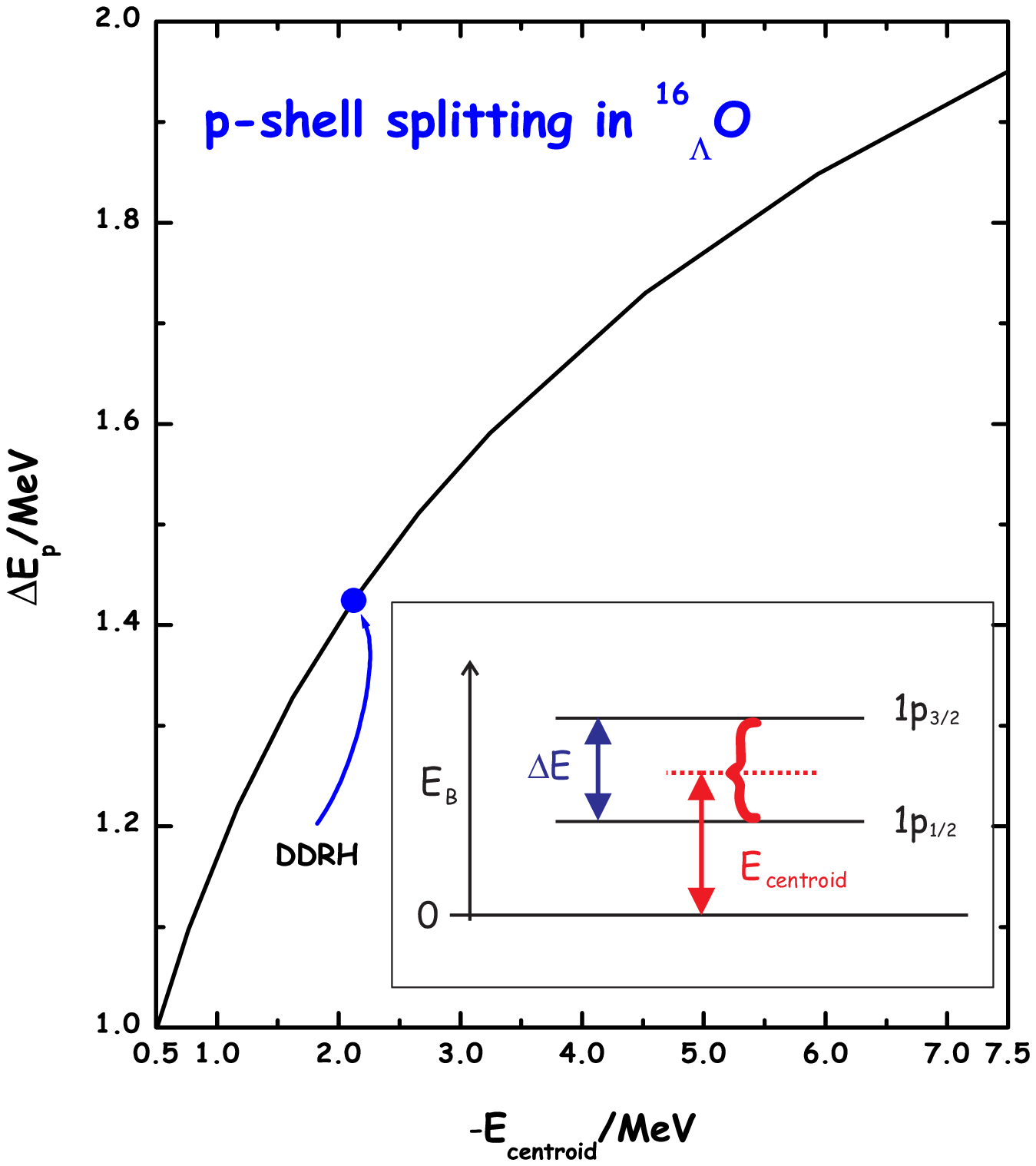}
\end{figure}
From weakly bound systems, e.g. neutron-rich dripline nuclei 
\cite{Lalazissis:1998if}, it is known that a s.o. doublet which approaches
its continuum threshold gets compressed before actually becoming unbound.
This is due to the so called avoided level crossing, reflecting the fact that a state
which becomes unbound has to cross other states lying lower in the continuum before
it reappears as a resonance. In a mean-field description this situation is in particular
found in the light p-shell hypernuclei. The effect is investigated in a model study by 
varying the $\omega\Lambda\Lambda$
coupling in $^{16}_\Lambda$O artificially such that the centroid energy E$_{centroid}$
of the 1p s.o. doublet approaches the continuum threshold (for notation see 
fig.~\ref{fig:so}). The resulting s.o. splitting $\Delta$E depends sensitively on the
location of E$_{centroid}$ relative to the continuum threshold. 
Therefore a clean extraction of the generic s.o. interaction 
from light hypernuclei is hardly to achieve because the threshold compression will always
be superimposed on the generic s.o. interaction.
However, around the $^{40}_\Lambda$Ca region the $\Lambda$ 1p doublet has moved far 
enough into the bound region such that the threshold effect ceases to be important.

\subsubsection{Wave function delocalization}
\begin{table}
\caption{Localization coefficients $N_{\Lambda,n}(r_{rms})$ as defined in 
         eq.~(\ref{eq:N}).}
\label{tab:deloc}
\begin{tabular}{l||rr|rr|rr|rr}
&\multicolumn{2}{c}{$^{40}_{\Lambda}$Ca}&\multicolumn{2}{c}{$^{51}_{\Lambda}$V}
&\multicolumn{2}{c}{$^{89}_{\Lambda}$Y}&\multicolumn{2}{c}{$^{208}_{\Lambda}$Pb} \\
&$N_\Lambda$&$N_n$&$N_\Lambda$&$N_n$&$N_\Lambda$&$N_n$&$N_\Lambda$&$N_n$\\
\hline
1p$_{3/2}$&0.51&0.71&0.59&0.75&0.68&0.81&0.79&0.88\\
1p$_{1/2}$&0.50&0.71&0.60&0.76&0.69&0.82&0.80&0.86\\
\hline
1d$_{5/2}$&0.23&0.47&0.33&0.54&0.47&0.64&0.66&0.80\\
1d$_{3/2}$&0.17&0.46&0.30&0.54&0.49&0.67&0.69&0.84\\
\hline
1f$_{7/2}$&--&--&--&--&0.27&0.45&0.52&0.72\\
1f$_{5/2}$&--&--&--&--&0.26&0.47&0.55&0.77\\
\end{tabular}
\end{table}
Another effect which diminnishes the influence of the s.o. interaction on the 
observed splitting by an additional reduction is the
delocalization of the $\Lambda$ wave function due to the shallow $\Lambda$ potential
(e.g. \cite{Rijken:1999yy}). This comes about because
the splitting is proportional to the overlap integral between the 
single particle density of the considered state and the s.o. potential which is well
localized in the nuclear surface. To determine the
importance of this delocalization we define the following measure for the fraction
of the $\Lambda$ and neutron wave function $F_{\Lambda,n}$ residing inside the 
nuclear rms radius: 
\begin{equation}
\label{eq:N}
N_{\Lambda,n}(R)=N_o\int_0^R dr\;r^2\;\left|F_{\Lambda,n}(r)\right|^2
\end{equation}
where $N_o$ is chosen such that $N_{\Lambda,n}(R)\rightarrow 1$ for $R\rightarrow\infty$
and $R=\sqrt{\left<r^2\right>}$. Results for different nuclei are
shown in tab.~\ref{tab:deloc}. It is obvious that this effect is especially pronounced
in light hypernuclei and in the weakly bound orbits in the heavy mass region.

We conclude from these investigations, that hyperon interactions still can be determined
provided that both the centroid energy and the relative splitting of a doublet are
considered. In this respect the new $\gamma$ spectroscopic results, reporting an almost
vanishing s.o. splitting in e.g. $^{13}_\Lambda$C \cite{Ajim}, point to an extremely weak
effective $\Lambda$ s.o. strength but for decisive conclusions on the generic interaction
more strongly bound heavy hypernuclei should be investigated.

\subsection{Hypernuclear Auger effect}
\label{sec:auger}
\begin{figure}
\label{fig:auger}
\caption{The partial life times of the 1g $\Lambda$ shell in $^{209}_\Lambda$Pb due to
         Auger neutron emission are plotted versus the kinetic energy of the outgoing
         neutron. The insert shows the $\Lambda$ and neutron level schemes for this
         transition where the involved states are marked.}
\includegraphics[height=.2\textheight]{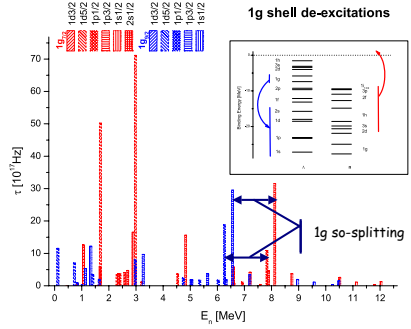}
\end{figure}
Auger rates for the capture of a $\Lambda$ and the de-excitation of heavy hypernuclei
are calculated within the DDRH model \cite{Keil:2000hk}. Different to a previous attempt
\cite{Likar:1986jj} our approach treats the binding mean-field, wave functions and 
interactions in a self-consistent manner.
The process is described by the production of a
neutron-particle--neutron-hole state due to the de-excitation 
of the $\Lambda$, where the particle
is unbound. A schematical illustration of the process for the 1g shell is shown in the
insert of fig.~\ref{fig:auger}.
The details of the calculation will be given elsewhere. Fig.~\ref{fig:auger}
shows the partial life times for the $\Lambda$ transitions starting from the 1g shell
plotted against the kinetic energy of the emitted neutron. The manifestation of the s.o.
splitting is marked. Due to the complexity of the full spectrum it will be important to
tag the $\Lambda$ initial state in the measurements of Auger spectra. A further analysis of
the global spectral features of the hypernuclear Auger effect is in progress.

\begin{theacknowledgments}
This work has been partially supported by DFG grant Le439/4-3 and the European 
Graduate School ``Complex Systems of Hadrons and Nuclei, Gie\ss en--Copenhagen''.
\end{theacknowledgments}


\begin{thebibliography}{99}

\bibitem{Keil:2000hk}
C.~M.~Keil, F.~Hofmann and H.~Lenske,
Phys.\ Rev.\ C {\bf 61}, 064309 (2000)
[nucl-th/9911014].

\bibitem{Likar:1986jj}
A.~Likar, M.~Rosina and B.~Povh,
Z.\ Phys.\ A {\bf 324}, 35 (1986).

\bibitem{JLabAuger} A. Margaryan {\it et al.}, JLab letter of intent (2000).

\bibitem{Lalazissis:1998if}
G.~A.~Lalazissis, D.~Vretenar, W.~Poschl and P.~Ring,
Nucl.\ Phys.\ A {\bf 632}, 363 (1998)
[nucl-th/9710013].

\bibitem{Rijken:1999yy}
T.~A.~Rijken, V.~G.~Stoks and Y.~Yamamoto,
Phys.\ Rev.\ C {\bf 59}, 21 (1999)
[nucl-th/9807082].

\bibitem{Ajim}
S.~Ajimura {\it et al.},
Phys.\ Rev.\ Lett {\bf 86}, 4255 (2001)


\end{thebibliography}


\end{document}